\documentclass[%
reprint,
superscriptaddress,
 amsmath,amssymb,
 aps,
prl,
]{revtex4-1}

\usepackage{notes2bib}
\usepackage{graphicx}
\usepackage{dcolumn}
\usepackage{bm}
\usepackage{color}
\usepackage{MnSymbol}

\definecolor{mygreen}{rgb}{0,0.85,0.45}
\definecolor{myred}{rgb}{1.0,0.44,0.41}
\definecolor{mypurple}{rgb}{0.57,0.49,0.84}
\definecolor{myblue}{rgb}{0,0.53,0.77}

\begin{document}

\preprint{APS/123-QED}

\title{Pattern formation during the impact of a partially-frozen binary droplet on a cold surface}

\author{Pallav Kant}
\affiliation{Physics of Fluids Group, Max Planck Center Twente for Complex Fluid Dynamics, University of Twente, 7500 AE Enschede, The Netherlands}
\email[]{p.kant@twente.nl}
\author{Henrik M{\"u}ller-Groeling}
\affiliation{Department of Physics and Astronomy, University of Heidelberg, Germany}
\author{Detlef Lohse}
\affiliation{Physics of Fluids Group, Max Planck Center Twente for Complex Fluid Dynamics, University of Twente, 7500 AE Enschede, The Netherlands}
\affiliation{Max Planck Institute for Dynamics and Self-Organization, Am Fa\ss berg 17, 37077 G{\"o}ttingen, Germany}
\email[]{d.lohse@twente.nl}


\date{\today}

\begin{abstract}

Impact of a droplet on an undercooled surface is a complex phenomenon as it simultaneously instigates several physical processes that cover a broad spectrum of transport phenomena and phase-transition.
Here, we report and explain an unexpected but highly relevant phenomenon of fingered growth of the solid-phase. 
It emerges during the impact of a binary droplet that freezes from the outside prior to the impact on the undercooled surface. We establish that the presence of pre-solidified material at the advancing contact line fundamentally changes the resulting dynamics, namely by modifying the local flow mobility that leads to an instability analogous to viscous fingering. 
Moreover, we delineate the interplay between the interfacial deformations of the impacting droplet and patterned growth of the solid-phase as disconnected patterns emerge at faster impacts.


\end{abstract}

\maketitle

Solidification of a droplet impacting on an undercooled surface is encountered in a wide range of technological applications, including material processing through splat quenching \cite{bennett1993splat}, high-resolution prototyping via additive manufacturing \cite{gao1994precise,waldvogel1997solidification, bhola1999parameters,liu2001high}, and ice accretion on aerodynamics surfaces \cite{bragg2007airfoil,shin1996characteristics}. 
In nearly all of these applications, the droplet liquid is not pure, but a mixture \cite{boley2014direct, ngo2018additive}. 
However, despite the practical importance,  the majority of investigations on this topic have been focused on the impact of a pure liquid droplet that simultaneously wets and solidifies upon its impact on an undercooled surface \cite{madejski1976solidification, pasandideh1998deposition, schremb2018normal, thievenaz2019solidification, gielen2020solidification, kant2020fast}.
Consequently, only little is known about the physico-chemical effects arising during the solidification of an impacting droplet with multiple components. 
In this Letter, we show that the overall freezing behaviour significantly alters if the impacting liquid is a {\it{binary}} mixture.
In particular, if one component is extremely volatile such that the resulting evaporative cooling is sufficient to freeze the impacting droplet from the outside prior to the impact.
Here, we adapt total-internal-reflection (TIR) imaging \cite{kant2020fast}, to explore the dynamics ensuing from the impact of a binary droplet on an undercooled surface.

In a typical experiment, we release a droplet of a binary mixture (80:20 \%volume) of hexadecane and diethyl ether of radius $R_\mathrm{drop} = 0.82 \pm 0.2$ mm, from the tip of a needle, such that it hits the horizontal surface of an undercooled sapphire prism with a vertical velocity $U$.
Both liquids are optically transparent at room temperature and completely miscible.
Hexadecane is non-volatile at room temperature, with a density $\rho = 770\,\mathrm{kg/m}^3$, surface tension $\sigma = 27 \,\mathrm{mN/m}$, dynamic viscosity $\mu = 3.47 \times 10^{-3}\,\mathrm{Pa~s}$.
On the contrary, at room temperature, diethyl ether is a highly volatile liquid with a low boiling point $36^\circ$C, and it has a high enthalpy of vaporization 27 kJ/mol.
Thus, while inflating a droplet of the binary mixture under ambient lab conditions the diethyl ether continuously evaporates.
As a consequence, the first pendant and then falling droplet cools down.
We find that this evaporative cooling is sufficient to freeze hexadecane at the droplet interface, which has a melting point $T_\mathrm{m} = 18^\circ$C.
Therefore, in our experiments, before the droplet impacts on the undercooled substrate, it already partially freezes from the outside.
For a detailed description of the experimental setup we refer to Supplementary Material.

\begin{figure*}
\centering
\includegraphics[clip, trim=0cm 0cm 0cm 0cm, width=0.95\textwidth]{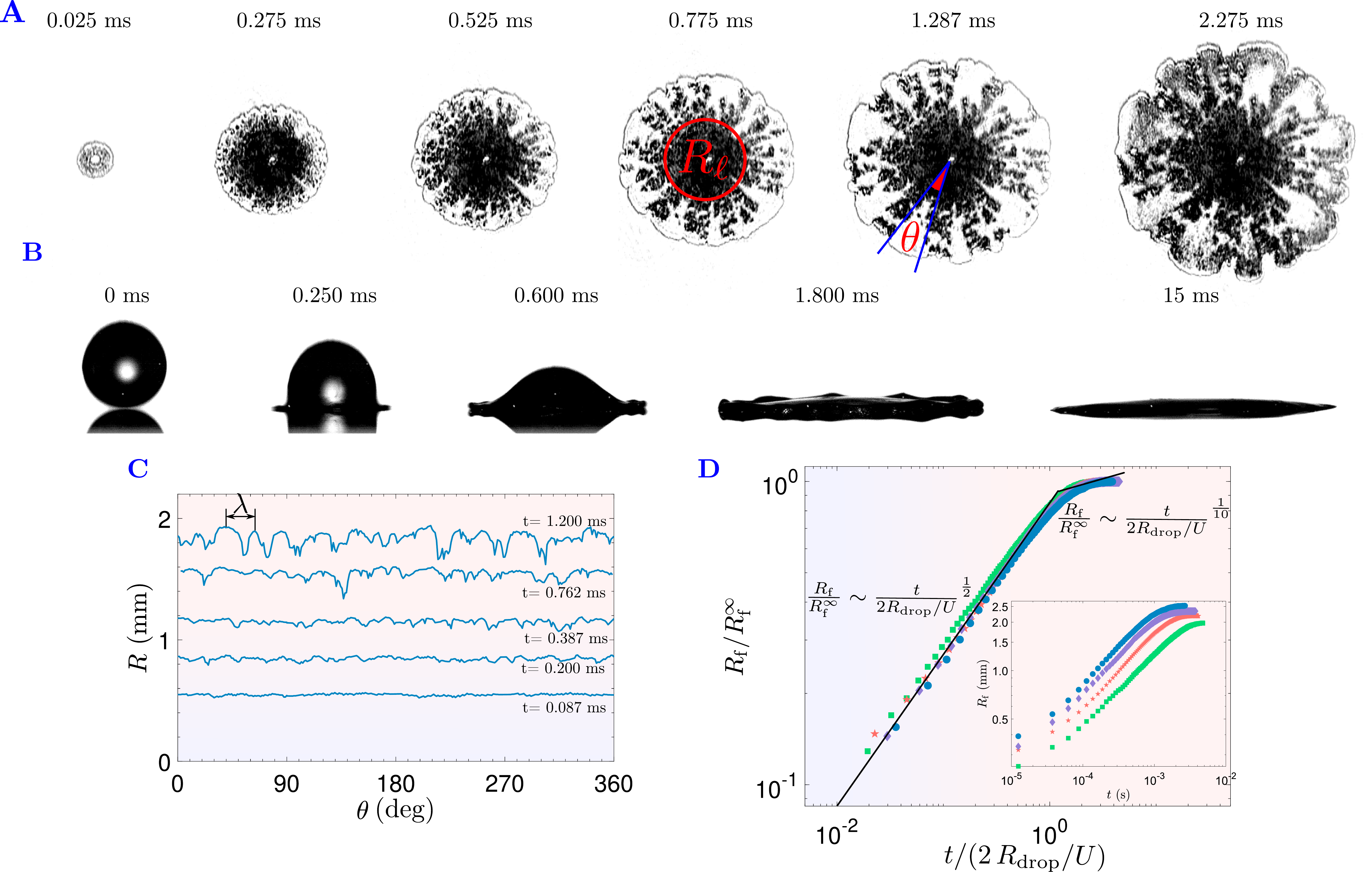}\\
\caption{\textbf{A}, Bottom-view snapshots, recorded \textsl{via} high-speed total-internal reflection imaging, highlighting the temporal growth of the solid-phase in finger-like patterns along the substrate after the impact of a partially-frozen droplet ($R_\mathrm{drop} = 0.85$ mm) of a binary mixture (80:20 \%vol) of hexadecane and diethyl ether. Here, $U =1.24$ m/s ($Re \sim 440$) and $T_\mathrm{s} = 6^\circ C$ as substrate temperature. Both the contact line and the solidified phase (hexadecane) are seen as grey-shaded regions in the TIR images. \textbf{B}, Side-view high-speed recordings: Typical deformations experienced by the partially-frozen interface as the droplet ($R_\mathrm{drop} = 0.82$) impacts on the substrate $T_\mathrm{s} = 6^\circ C$ with $U = 1.45$ m/s ($Re \sim 517$). \textbf{C}, Temporal evolution of the azimuthal undulations along the radius $R(\theta)$ of the contact line during the expansion of the wetted area shown in sequence in \textbf{A}. \textbf{D}, Scaled master curve of the spreading radius $R_\mathrm{f}/R_\mathrm{f}^{\infty}$ versus time $t/(2 R_\mathrm{drop}/U)$ for droplet impacts on the undercooled substrate, $T_\mathrm{s} = 6^\circ$C, at different impact velocities $U$ = (\textcolor{mygreen}{$\filledmedsquare$}) 0.88 m/s (\textcolor{myred}{$\filledlargestar$}) 1.45 m/s (\textcolor{mypurple}{$\filledmedlozenge$}) 1.90 m/s (\textcolor{myblue}{$\bullet$}) 2.25 m/s. Inset: Raw data for the temporal evolution of $R_\mathrm{f}$ for these velocities. }
\label{fig:fig1}
\end{figure*}

\begin{figure*}[htb]
\centering
\includegraphics[clip, trim=0cm 0cm 0cm 0cm, width=0.95\textwidth]{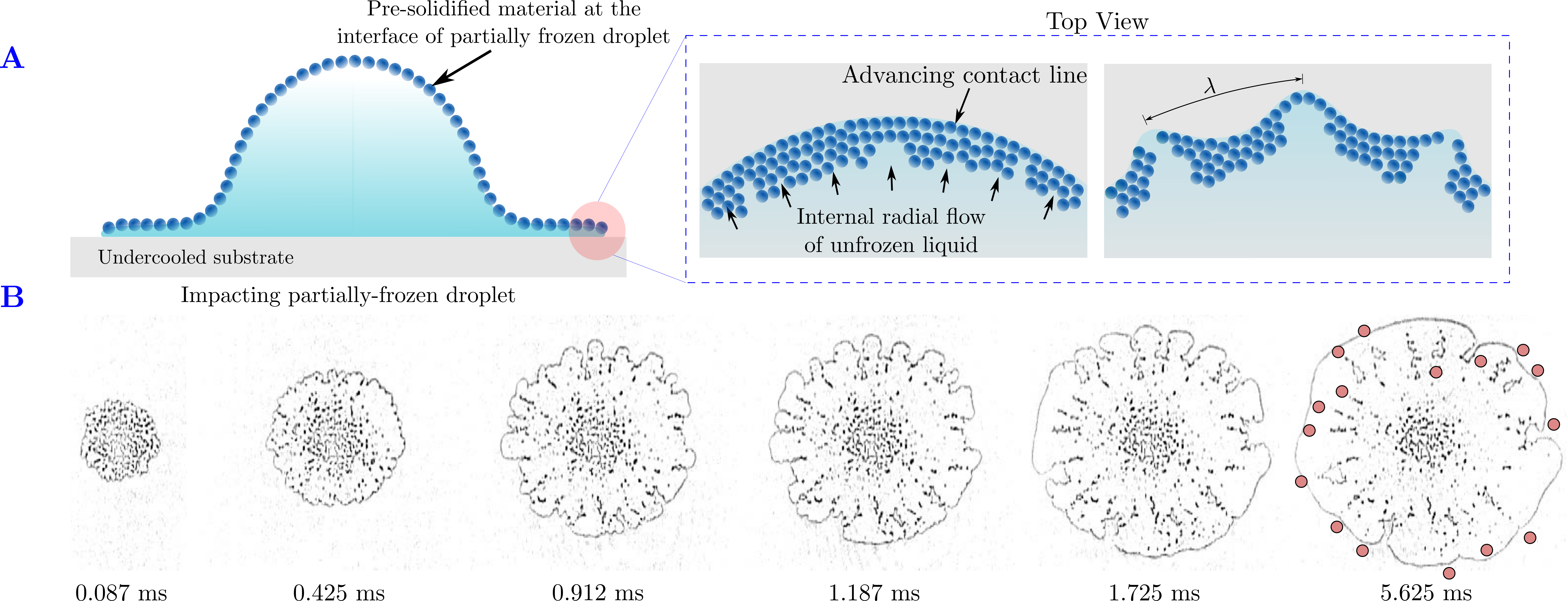}\\
\caption{\textbf{A}, Sequence of events leading to the destabilization of the advancing contact line and fingered growth of the solid-phase are outlined schematically. While a partially-frozen droplet deforms on the substrate, pre-solidified material accumulates at the advancing contact line. This results in a mobility contrast between the unfrozen fluid away from the contact line and its vicinity, thus viscous-fingering occurs. \textbf{B}, A liquid marble is used as a model system to mimic a droplet that is frozen from the outside. The bottom-view sequence recorded \textsl{via} TIR shows a typical outcome of the impact of the liquid marble ($R_\mathrm{drop} = 1.06$ mm, $U = 1.45$ m/s). Notably the emergent features, namely destabilization of the contact line and clustering of particles in radially growing finger-like patterns, closely resemble the patterns observed for the impact of a partially-frozen droplet. Small circles at the periphery of the droplet footprint highlight the fingered clusters of particles.}
\label{fig:fig2}
\end{figure*}

The TIR imaging (Fig.\,\ref{fig:fig1}a) reveals that the impact of the partially-frozen binary droplet on the undercooled surface unexpectedly results in the growth of the solid phase in finger-like patterns along the substrate.
These are reminiscent of classical viscous-fingering patterns in a radial geometry \cite{homsy1987viscous}. 
We observe that just before the droplet makes contact with the substrate, its crystal-laden interface deforms to entrap an air-bubble \cite{driscoll2011ultrafast,bouwhuis2012maximal}.
The footprint of this entrapped bubble is visible in the middle of the wetted area in all the snapshots in Fig.\,1a.
Upon establishing the contact with the substrate, the droplet footprint first expands uniformly in all directions and the wetted area solidifies uniformly.
However, soon after the touch-down (for $t > 0.125$ ms), the expanding wetted area loses axisymmetry and shallow azimuthal undulations of small wavelength ($\lambda \sim 100 \mu$m) appear at the contact line.
At later times these undulations at the contact line gradually grow into deeper structures with larger wavelength ($\sim 1$ mm) until the droplet stops spreading at $t \sim t_\mathrm{I} = 2 R_\mathrm{drop}/U$; see Fig.\,\ref{fig:fig1}c.
This growing instability at the contact line manifests itself through fingered growth of the solid-phase in the region $R > R_\ell$; $R_\ell$ indicates the radius of the uniformly solidified area in the middle of the droplet footprint \footnote{The whole wetted area eventually solidifies at times $t > t_\mathrm{I}$. and is controlled by the substrate temperature $T_\mathrm{s}$.}.
Correspondingly, the instant $t_\ell$, defined by the time when the footprint radius reaches $R_\ell$, marks the onset of the fingered growth of the solidified-phase.
Note that the high-speed imaging of droplet spreading reveals that the fingered growth of the solid phase, even though it appears continuous, is in fact a result of continuous tip-splitting \cite{paterson1981radial, homsy1987viscous} and merging of azimuthal undulations at the contact line \footnote{Merging of the frontal undulations is a peculiar feature of our system, owing to the highly wetting nature of the substrate, with static contact angle less than $10^\circ$, that promotes the spreading of the impacting liquid in all directions.}.


Remarkably, the overall impact dynamics remains unaffected by the presence of solidified material at the droplet interface \footnote{See supplementary material}. 
Immediately after the impact, a thin-lamella emerges from the base of the partially-frozen droplet (Fig.\,\ref{fig:fig1}b), as a flux of momentum is directed tangentially along the substrate.
This lamella expands non-axisymmetrically along the substrate, owing to the undulations at the contact line.
Strikingly, despite the persistent azimuthal perturbations at the contact line, at early times the temporal evolution of the radius of the best circular fit to the droplet footprint $R_\mathrm{f}$ (Fig.\,\ref{fig:fig1}d) closely follows the standard power-law like behaviour $R_\mathrm{f} \propto t^{\frac{1}{2}}$ \cite{riboux2014experiments}.
At later times, as the impacting droplet continues to spread into a thin-layer viscous effects slow down its spreading rate. Consequently, the spreading beahviour gradually transitions into Tanner's regime $R_\mathrm{f} \propto t^{\frac{1}{10}}$ for a brief duration before the droplet footprint reaches its equilibrium position.
Notably, we are able to collapse the spreading curves obtained for different impact velocities by re-scaling the spreading radii with the maximum radius $R_\mathrm{f}^\infty$ and the time with characteristic timescale $t_\mathrm{I}$.

We stress that the nature of contact line instability observed in our experiments is drastically different from the fingering instability \cite{allen1975role, thoroddsen1998evolution} reported for iso-thermal drop-impacts at vey high Reynolds number, $Re = 2 \rho\,U R_\mathrm{drop}/\mu \sim 10\,000$.
Thoroddsen and Sakakibara \cite{thoroddsen1998evolution} suggested that during fast impacts right before the impacting droplet makes contact with the substrate, it feels the presence of the solid surface and strongly decelerates in the air.
This develops the seeds for the azimuthal instability by a mechanism similar to the Rayleigh-Taylor instability.
Thus instantly after a rapidly decelerated droplet makes contact with the substrate azimuthal undulations appear at the contact line.
Most importantly, their experiments indicated that the fundamental wavelength $\lambda^{\mathrm{RT}}$ of these undulations hardly varies as the droplet spreads outward, and is given by:
\begin{equation}
\lambda^{\small{\mathrm{RT}}} = \sqrt{\frac{3\,\sigma}{a^{*} \rho}}.
\end{equation}
Here, $a^{*}$ is the initial deceleration experienced by the droplet and $\sigma$ the surface tension of the impacting liquid.
On the contrary, in our experiments ($Re < 1000$), no perturbations are observed at the contact line immediately after the droplet touch down.
Moreover, as seen in Fig.\,\ref{fig:fig1}a, the angular displacement between two consecutive fingers remains unchanged while the droplet footprint expands. 
Therefore the dominant wavelength of the perturbations at the contact lines changes in time (Fig.\,1c).
This suggests that the contact line instability arising from the impact of a partially-frozen droplet does not originate from the strong deceleration experienced by droplet interface during the impact.

The physical mechanism responsible for the destabilization of the advancing contact line and corresponding fingered growth of the solid phase along the substrate is outlined schematically in Fig.\,\ref{fig:fig2}a.
We hypothesize that upon impact, while a partially-frozen droplet deforms and spreads into a thin pancake along the substrate, the pre-solidified material at its interface accumulates in the vicinity of the advancing contact line, which locally enhances the flow resistance.
This results in a mobility contrast between the unfrozen fluid away from the contact line and its vicinity, leading to particle-induced viscous-fingering \cite{tang2000stability, kim2017formation}.
The resultant inhomogeneous distribution of pre-solidified material causes interfacial deformations and fingered growth of the solid-phase along the substrate.

To confirm the above hypothesis, we performed iso-thermal impact experiments with particle-coated drops \textsl{i.e.}, so called liquid marbles \cite{aussillous2001liquid, aussillous2006properties}, which closely mimic droplets that are frozen from the outside.
Strikingly, the outcome of these experiments exhibits all the essential features we have seen for the above described partially-frozen binary droplet impacts, \textsl{i.e.} destabilization of the contact line and clustering of particles in finger-like patterns along the substrate (Fig.\,\ref{fig:fig2}b).
Furthermore, the non-linear evolution of the frontal undulations through tip-splitting and merging events corroborates our hypothesis.
This also reflects that the extent of viscous-fingering during the impact of a partially-frozen binary droplet crucially depends on the local concentration of the pre-solidified material at the contact line as it controls the mobility contrast between the unfrozen fluid away from the contact line and its vicinity \cite{kim2017formation}.
Correspondingly, for droplets with lower volume fraction (less than 10\%) of volatile diethyl ether, no fingered-growth of the solid-phase and unstable motion of the contact line is observed (Movie-3) because of the limited evaporative-freezing of such droplets.

\begin{figure}
\centering
\includegraphics[clip, trim=0cm 0cm 0cm 0cm, width=0.4\textwidth]{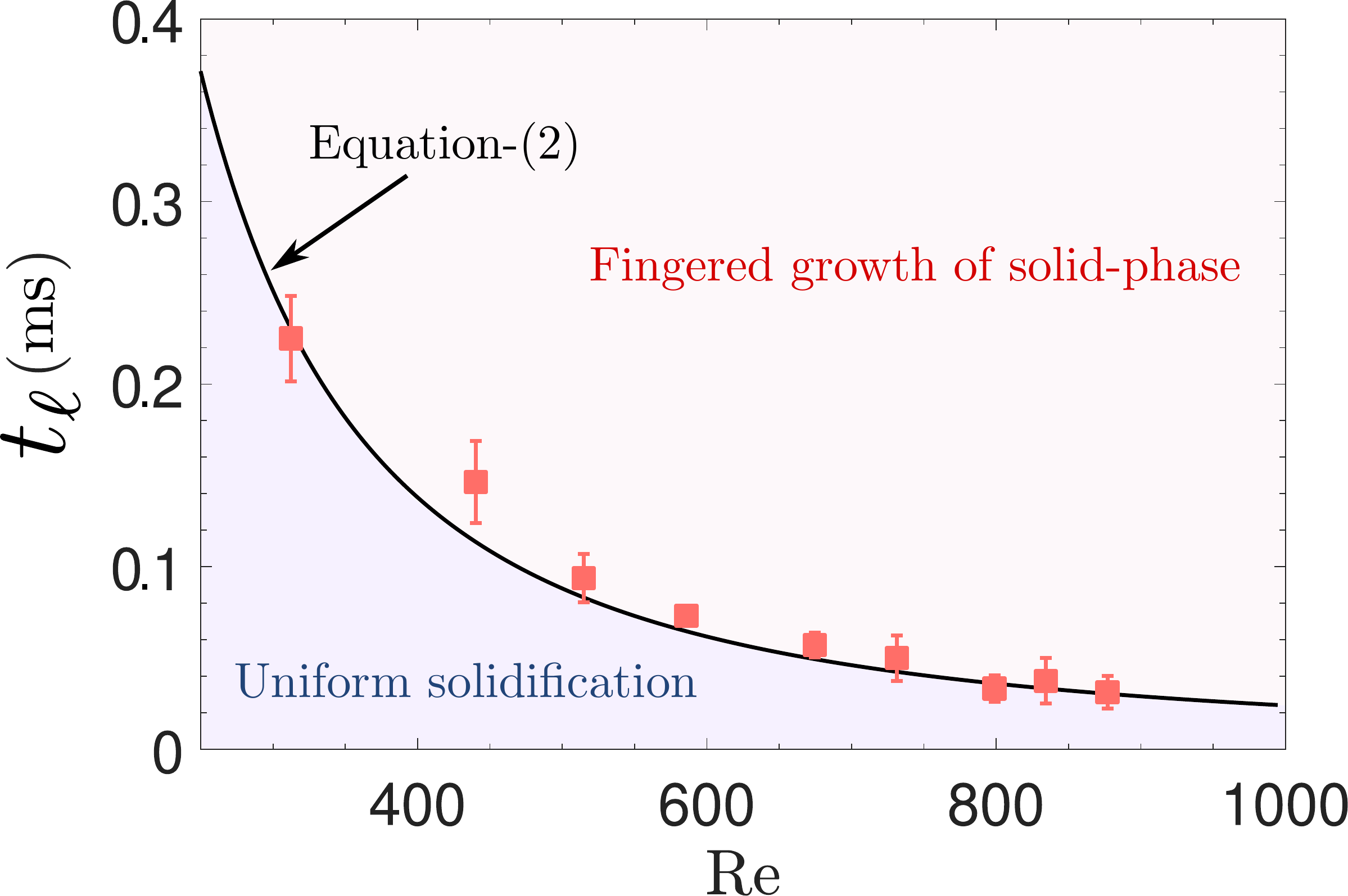}\\
\caption{A comparison between the experimentally measured (red symbols) onset $t_\ell$ of the fingered growth of the solid-phase and the theoretical predictions of the instant of lamella ejection, computed from equation (2) (solid black-line). The experimental values correspond to the impacts of partially-frozen droplets of 80:20 \%vol mixture of hexadecane and diethyl ether on the undercooled substrate at $T_\mathrm{s} = 9^\circ$C.}
\label{fig:fig3}
\end{figure}

To develop a quantitative understanding of the patterns arising from the impact of partially-frozen droplets, we now estimate the onset of fingered-growth of the solid-phase $t_\ell$.
It has been demonstrated both experimentally \cite{chevalier2006inertial} and theoretically \cite{dias2011influence} that inertial effects stabilize the non-uniform motion of an interface during the viscous-fingering instability.
Likewise, in our experiments, the fingered growth of the solid-phase does not emerge immediately after the impact of a droplet when the inertial effects still dictate the spreading, but only some time later.
This suggests that the emergence of the fingered patterns of the solid-phase is a direct consequence of a crossover in the dominant physical forces during the motion of the contact line after impact, namely the crossover to dominance of the viscous forces.
It must be noted that this crossover between the dominant physical forces is also closely related to the ejection of the lamella at the base of the droplet after the touch-down \cite{mongruel2009early}.
The key point here is that the local Reynolds number within the ejected lamella is small thanks to its small thickness ($50 - 100\,\mu$m), that viscous effects significantly influence the subsequent dynamics.
Thus, the instant of lamella ejection marks the inception of the spreading regime where viscous effects begin to play an important role.
Accordingly, we use it as an estimate for the onset $t_\ell$ of the fingered patterns.
Since the presence of solidified material at the droplet interface hardly changes the overall droplet deformation during the impact (Fig.\,\ref{fig:fig1}b), to compute $t_\ell$,  we use the formulation proposed by Riboux and Gordillo \cite{riboux2014experiments}, namely,
\begin{equation}
\sqrt{3}/2 Re^{-1} t_\ell^{-1/2} + Re^{-2} Oh^{-2} = c\,t_\ell^{3/2}.\\
\label{eq:eq1}
\end{equation}
Here, $Oh = \mu/\sqrt{2 \rho\,R_\mathrm{drop}\,\sigma}$ is the Ohnesorge number and $c = 0.8$ is a numerical constant that accounts for the proportionality constant in the scaling for lamella thickness; see Ref \cite{riboux2014experiments}.
Note that equation (2) is a direct consequence of the condition that the fluid particles close to the substrate move faster than the contact line.
A comparison between the experimental measurements of the crossover time $t_\ell$ marking the instant when the contact line reaches $R_\ell$ and the time of ejection of the lamella determined from equation (\ref{eq:eq1}) is shown in Fig.\,\ref{fig:fig3}.
The close agreement between the two confirms that the fingered growth of the solid-phase is promoted by the viscous effects within the expanding lamella.

\begin{figure}
\centering
\includegraphics[clip, trim=0cm 0cm 0cm 0cm, width=0.48\textwidth]{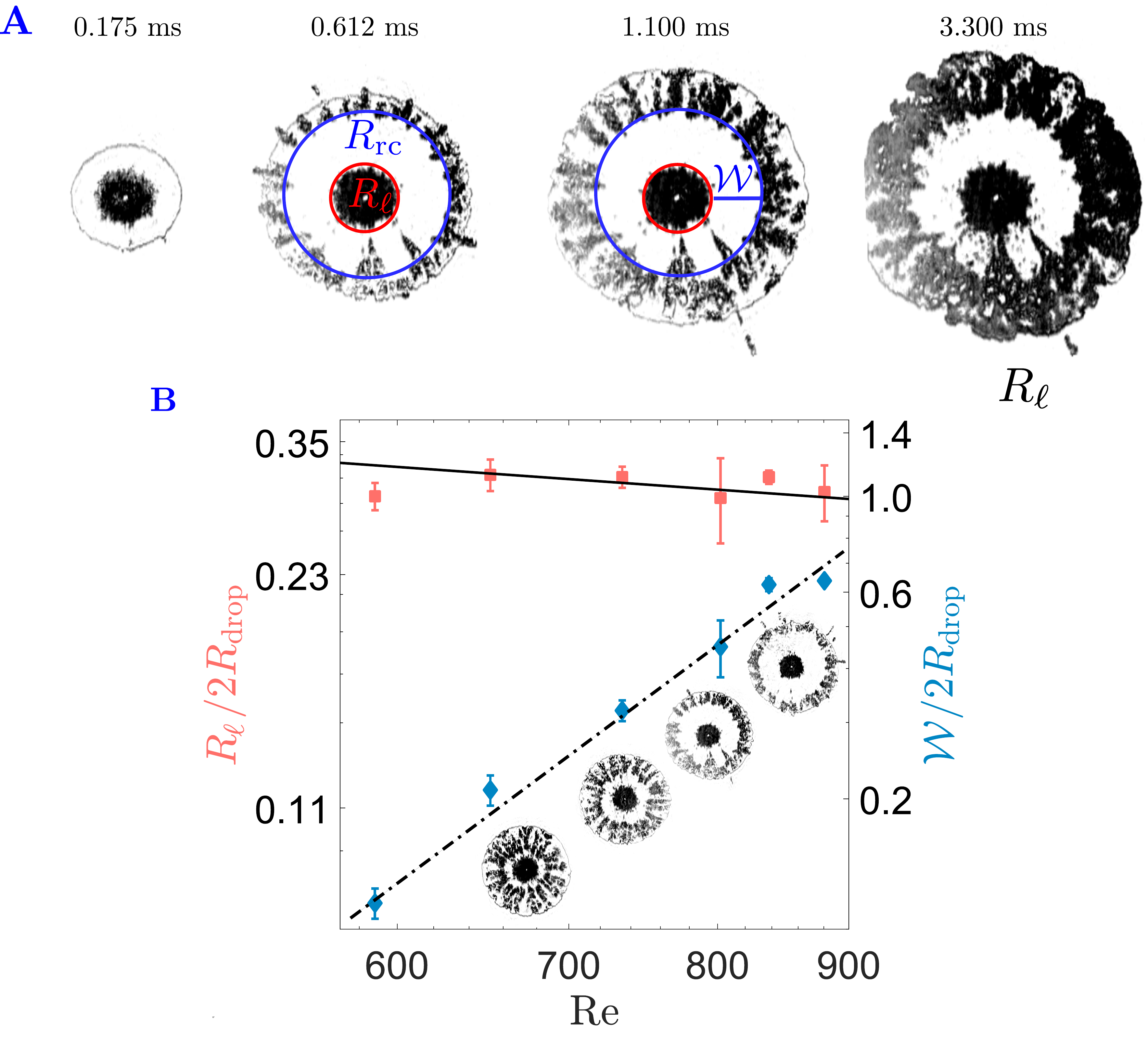}\\
\caption{\textbf{A}, Temporal growth of the solid-phase in disconnected patterns after the impact ($U = 2.25$ m/s, $Re \sim 800$) of a binary droplet of a mixture (80:20 \%vol) of hexadecane and diethyl ether on an undercooled substrate, $T_\mathrm{s} = 9^\circ$C.  \textbf{B}, At higher impact velocities ($Re > 550$), the front of the ejected lamella initially lifts-off from the substrate at the location $R_\ell$.  Consequently no particle-induced viscous-fingering takes place until the lamella re-wets the substrate at $R_\mathrm{rc}$. The extent of uniformly solidified central patch $R_\ell$ marks the location of lamella ejection. For increasing impact velocities, $R_\ell$ slowly decreases as $R_\ell/2 R_\mathrm{drop} \sim Re^{-1/4}$ (solid-black line). Beside, the width of the unfrozen annular gap $\mathcal{W} = R_\mathrm{rc} - R_\ell$ widens for increasing $U \sim Re$. The dashed-line is a guide to the eyes. The experimental data plotted here corresponds to the impacts of partially-frozen droplets of 80:20 \%vol mixture of hexadecane and diethyl ether on the undercooled substrate at $T_\mathrm{s} = 9^\circ$C.}
\label{fig:fig4}
\end{figure}

In addition, we find that the kinematics of the ejected lamella significantly affects the freezing morphologies.
In our experiments, for higher impact velocities $Re > 550$, the front of the lamella initially dewets the solid surface as it lifts-off from the substrate. 
Only later it re-contacts the substrate (Movie-4) due to the capillary retraction \cite{riboux2014experiments}.
Note that the lamella lift-off was observed for slightly higher values of $Re> 700$ for iso-thermal impacts.
This upward motion of the lamella, however, for a brief period, leads to the disconnected fingered growth of the solid-phase at early times; see Fig.\,\ref{fig:fig4}a (Movie-5).
We observe that upon the lift-off of the lamella at location $R_\ell$ no fingered growth of the solid phase occurs within the annular region of width $\mathcal{W} = R_\mathrm{rc} - R_\ell$; $R_\mathrm{rc}$ is the location where lamella re-contacts the substrate.
This indicates that during the lamella lifts-off from the substrate no accumulation of pre-solidified material occurs at the contact line. 
Consequently, no particle induced viscous-fingering follows and the droplet footprint expands uniformly until the lamella re-wets the substrate.
The width of this unfrozen annular region remains unchanged while the droplet continues to spread, however, it solidifies uniformly at times $t > t_\mathrm{I}$.
Further, as seen in Fig.\,\ref{fig:fig4}b, in our experiments, the size of the central patch nearly remains constant for increasing impact velocity $U$.
This behaviour closely resembles the dynamics of lamella ejection in an iso-thermal scenario, for which in the high-$Oh$ limit it can be shown that $R_\ell$ decreases slowly for increasing impact velocity as $R_\ell/2 R_\mathrm{drop} \sim Re^{-1/4}$ \cite{note}.
On the other hand, we note that the width of the unfrozen annular gap, widens monotonically (linear relation $\mathcal{W}/2 R_\mathrm{drop} \propto Re$) for increasing impact velocities, see Fig.\,\ref{fig:fig4}b.
This is expected as the tangential momentum imparted to the fluid particles entering the lamella increases with $U$.
However, to quantitatively understand this behaviour a detailed modelling of the ballistic motion of the lamella is required, which is beyond the scope of this work.

Overall, we have analysed freezing of impacting binary droplets, unveiling the strong influence of the multi-component nature of the droplet and both evaporation and freezing, in addition to the already known effects of substrate undercooling and impact velocity.
We show that the evaporative-freezing of the droplet prior to the impact fundamentally alters the overall freezing behaviour.
The observed contact line instability and fingered growth of the solid-phase is explained as a consequence of particle-induced viscous-fingering arising from the accumulation of solidified material in the vicinity of the advancing contact line.
Further, we identify the role of both the kinematics and the dynamics of interfacial deformations experienced by the droplet at early-times on the resultant freezing patterns.

Most importantly, our results offer a new perspective to understand the freezing of impacting droplets.
The insights gained form these investigations can further be generalized to scenarios where impacting droplets already undergo phase-transition prior to the impact on the substrate, for instance a partially-frozen rain droplet hitting an aerodynamic surface, a topic of great interest and utility.
We expect that our findings might inspire further studies addressing the complexities associated with the freezing of multi-component droplets.

\section*{acknowledgments}
The authors thank Pierre Chantelot for valuable discussions, Valentina Dondei for her help in liquid marble impact experiments and Gert-Wim for the technical support in building the experimental setup.  We acknowledge the funding by Max Planck Center Twente, NWO and from the ERC Adv. Grant DDD 740479.

\bibliography{bibliography}

\begin{thebibliography}{33}
\bibitem [{not()}]{note}%
  \BibitemOpen
  \href@noop {} {\bibinfo  {journal} {The distance traversed before the
  ejection of the lamella can be computed as, $R_\ell = \int_0^{t_\ell} \dot
  R\,d\mathrm{t}$. Here, $\dot{R}$ is the contact line velocity immediately
  after the impact that scales as $\dot{R} \sim t^{-1/2}$ (Fig.\,1d). In the
  high-$Oh$ limit, the time of lamella ejection can be estimated as
  $t_\ell/(R_\mathrm{drop}/U) \sim Re^{-1/2}$, from equation (2). Thus, we find
  the scaling $R_\ell/R_\mathrm{drop} \sim Re^{-1/4}$}\ }\BibitemShut {NoStop}%
\end{thebibliography}

\end{document}